# Reduction of thermal conductance by coherent phonon scattering in two-dimensional phononic crystals of different lattice types

Roman Anufriev[1] and Masahiro Nomura[1,2,*]

[1]*Institute of Industrial Science, University of Tokyo, Tokyo, 153-8505, Japan*

[2]*Institute for Nano Quantum Information Electronics, University of Tokyo, Tokyo, 153–8505, Japan*

*Email address: nomura@iis.u-tokyo.ac.jp

The impact of lattice type, period, porosity and thickness of two-dimensional silicon phononic crystals on the reduction of thermal conductance by coherent modification of phonon dispersion is investigated using the theory of elasticity and finite element method. Increase in the period and porosity of the phononic crystal affects the group velocity and phonon density of states and, as a consequence, reduces the in-plane thermal conductance of the structure as compared to unpatterned membrane. This reduction does not depend significantly on the lattice type and thickness of phononic crystals. Moreover, the reduction is strongly temperature dependent and strengthens as the temperature is increased.

**I. INTRODUCTION**

Phononic nanostructures are regarded as possible candidates for applications involving phonon management [1–3], due to the possibility to manipulate the flux of vibrational energy (i.e., heat or sound) by band engineering. Indeed, on the one hand, phononic crystals may exhibit complete phononic bandgaps, i.e., regions of frequency where the propagation of elastic waves is forbidden in any direction; the physics of this phenomenon has been investigated both theoretically [3–9] and experimentally [5,10] in various types of two-dimensional (2D) phononic crystals. On the other hand, Bragg diffraction and local resonances in the periodic media result in the flattening of branches in phonon dispersion in a wide range of frequencies. This change of phonon dispersion leads to changes of the group velocities of phonons [11–13] and the density of states (DOS) [11,13–15]. These effects originate from the wave nature of phonons and are associated with coherent scattering, which is the process when phonons preserve their phase after a scattering event, in contrast to various incoherent scattering processes when phonons do not preserve their phase. Such modifications of phonon dispersion are thought to have the ability to reduce the thermal conductivity of nanostructures, in addition to reduction by incoherent scattering mechanisms such as surface [16–18], impurity [19], and Umklapp [17,20] scattering processes. Recent experiments demonstrated very low values of thermal conductivity at room temperature in freestanding thin-film silicon nanostructures with 2D square [14,21] and hexagonal [22] arrays of holes. To explain these low values of thermal conductivity, Dechaumphai *et al.* [12] developed a model that takes into account both coherent and incoherent scattering mechanisms, and Lacatena *et al.* [23] used a molecular dynamic approach, whereas Jain *et al.* [24] and Ravichandran *et al.* [25] argued that coherent scattering is unlikely to appear at room temperature and some of these experimental results could be explained by treating phonons as particles with bulk properties. However, Zen *et al.* [11] demonstrated that, in 2D phononic structures, at sub-kelvin temperatures, thermal conductance is totally controlled by coherent modifications of phonon dispersion, while incoherent mechanisms play a negligible role. An overview of recent theoretical and experimental investigations shows that coherent scattering of phonons can play an important role in heat transport, while its dependence on the geometry and temperature of the structure remains a disputable matter. No systematic theoretical studies of this effect are available in the literature.

In this study, we use the theory of elasticity and finite element method (FEM) to investigate theoretically the impact of the phononic structure design on the reduction of the group velocity and DOS in the in-plane directions. We demonstrate the dependence of thermal conductance reduction on the period, thickness, radius-to-period ratio, and lattice type of the phononic crystal structure and discuss the temperature dependence of this effect.

## II. SIMULATION OF PHONONIC CRYSTALS

We simulate infinite periodic arrays of holes in thin freestanding silicon membranes [Fig. 1(a)] with various periods ($a$), thicknesses ($h$) and hole radii ($r$). An infinite structure is simulated by considering a three-dimensional unit cell of finite thickness with Floquet periodic boundary conditions applied in the $x$–$y$ plane [6]. At the low temperature limit, the wavelengths of phonons are longer than the atomic scale, so we can use classical elasticity theory to compute the phonon modes. We use FEM, implemented by COMSOL MULTIPHYSICS® v4.4 software, to calculate numerically the phonon dispersion $\omega(k)$ from the elastodynamic wave equation:

$$\mu \nabla^2 \boldsymbol{u} + (\mu + \lambda) \nabla (\nabla \cdot \boldsymbol{u}) = -\rho \omega^2 \boldsymbol{u}, \tag{1}$$

with $\boldsymbol{u}$ as the displacement vector, $\rho = 2329$ kg m$^{-3}$ as the mass density, and $\lambda = 84.5$ GPa and $\mu = 66.4$ GPa as the Lamé parameters of silicon. First we calculate the eigenfrequencies for the wave vectors at the periphery of the irreducible triangle of the first Brillouin zone (BZ) [Fig. 1(b)], and then we evaluate the eigenfrequencies in the interior of the first BZ as an extrapolation of the values at the periphery within the triangle [Fig. 1.(c)] [26]. As a reference structure we use an unpatterned membrane of the same thickness, in which case the dispersion is obtained from analytic Rayleigh–Lamb equations [27]. To study the in-plane heat transport through the structure, we calculate the heat flux spectra $Q(\omega, T)$ at a given temperature ($T$) as:

$$Q(\omega,T) \propto \sum_m \int_0^{FBZ} \hbar\omega_m |\vec{v}_m(k)| f(\omega_m, T) d\vec{k} \qquad (2)$$

where $v_m$ and $D_m$ are the group velocity and the DOS calculated with close attention to the band intersections, and $f$ is the Bose-Einstein distribution [28]. The integrals are evaluated over the entire first BZ for each mode ($m$) and then summed. Figures 1(d), 1(e), and 1(f) show typical spectra of the average group velocity, DOS, and heat flux calculated from the obtained band diagram. In this work the average group velocity at a given frequency is calculated as an average between group velocity values obtained for different modes and in all in-plane directions at this frequency.

To verify the validity of our calculation, we simulated the same structures as those studied in the literature [4,6,12,29] and found obtained band diagrams and group velocity spectra to be in agreement with the literature. Moreover, we found good agreement with heat flux spectra and thermal conductivity results reported by Maasilta *et al.* [11,30], which proves that extrapolation of the eigenfrequencies in the interior of the first BZ and imperfection of our band sorting algorithm do not produce significant inaccuracy.

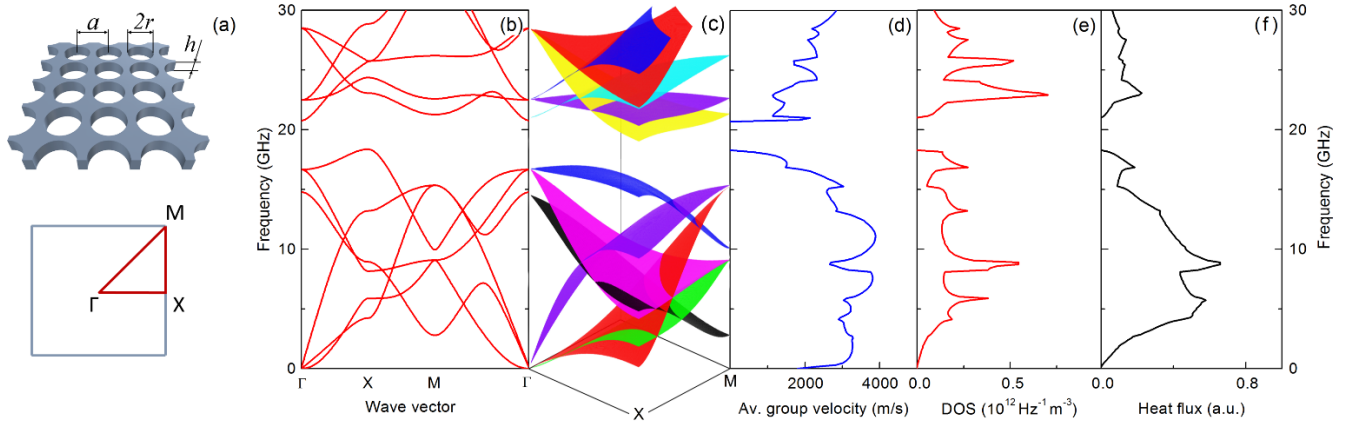

FIG. 1. (Color online) (a) Scheme of the simulated structure, its unit cell, and the first BZ with high symmetry points: Γ (0, 0), X (π/a, 0), and M (π/a, π/a). (b) Unsorted band diagram of the structure with $a = 160$ nm, $r/a = 0.45$, and $h = 80$ nm plotted at the high symmetry points. (c) Sorted band diagram plotted in the interior of the irreducible triangle of the first BZ. Spectra are shown for (d) the average group velocity, (e) DOS, and (f) heat flux.

## III. IMPACT OF THE DESIGN

In this work we mainly study phononic crystals of the same dimensions and lattice types as those studied experimentally in our recent work [31], i.e. 80-nm-thick square, hexagonal [Fig. 2(a)], and honeycomb [Fig. 2(b)] lattice structures with periods of few hundreds of nanometers. Figures 2(c) and 2(d) show the corresponding phonon dispersions of hexagonal and honeycomb lattices, respectively.

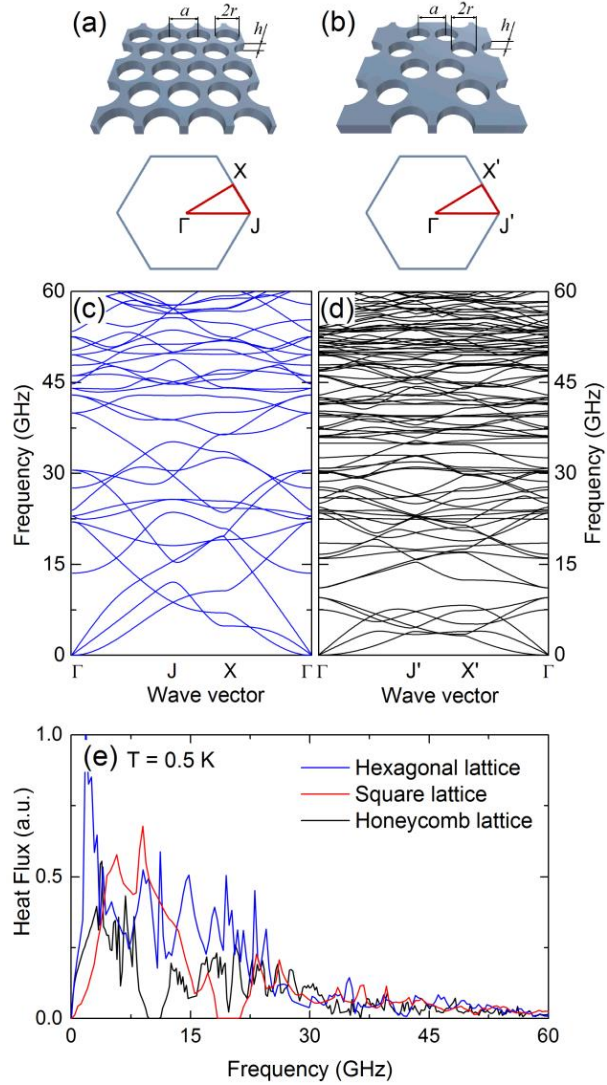

FIG. 2. (Color online) Schemes of the structures and first BZs with high symmetry points for (a) hexagonal lattice: Γ (0, 0), J ($4\pi/3a$, 0), and X ($\pi/a$, $\pi/\sqrt{3}a$), and (b) honeycomb lattices: Γ (0, 0), J' ($4\pi/3\sqrt{3}a$, 0), and X' ($\pi/\sqrt{3}a$, $\pi/\sqrt{3}a$). The band diagrams for (c) hexagonal and (d) honeycomb structures, and (e) the heat flux spectra in square, hexagonal and honeycomb structures at 0.5 K ($r/a = 0.45$, $a = 160$ nm, and $h = 80$ nm).

Despite the similarities between the lattices, the band diagrams are completely different, yet the corresponding spectra of the heat flux are quite similar [Fig. 2(e)], which shows that density of bands in band diagrams does not reflect heat transport properties. Analyzing the appearance of the phononic bandgap in the different lattices, we found the same trends as those known from the literature [4,8,9,11,29,32]: the width of the bandgap increases as a function of the *r/a* ratio, while, as a function of the *h/a* ratio, the widest bandgaps are expected at around the values of 0.5 and 1. However, the bandgap alone may play only a limited role in the suppression of thermal conductance at temperatures above the sub-kelvin range, as is evident from the heat flux spectra [Fig. 2(e)]: the region covered by the bandgap is rather small even at 0.5 K, as compared to the whole range of frequencies, and structures with and without a bandgap do not exhibit a significant difference in thermal conductance reduction, as we will see in the following sections. For this reason, we focus our study on the changes in the group velocity and DOS.

To illustrate how coherent modifications of phonon dispersion impact heat transport, we consider two square lattice nanostructures with the same thickness (80 nm) and *r/a* ratio (0.4), but different periods (80 and 320 nm). Figure 3(a) demonstrates that both phononic structures show significant reduction of the group velocity, as compared to the unpatterned membrane, but this reduction is larger in the structure with a longer period ($a = 320$ nm). This reduction is a direct consequence of band flattening in phononic crystals [11,12,33]. The DOS of the phononic structures are similar to that of the membrane. Yet again, the structure with $a = 320$ nm demonstrates a lower DOS, while DOS of the structure with $a = 80$ nm even exceeds that of the membrane in the low frequency part. [Fig. 3(b)]. As a consequence, the heat flux, which is proportional to the product of the group velocity and DOS, shows a reduction for both phononic structures as compared to the membrane, except for the low frequency part [13] [Fig. 3(c)]. This reduction is stronger in the structure with a longer period, due to the lower DOS and slightly lower group velocity spectra. The same reduction is observed also in hexagonal and honeycomb lattices. This result is consistent with recent experimental and theoretical data where the heat flux in SiN phononic structures was also reduced significantly as the period was increased [11].

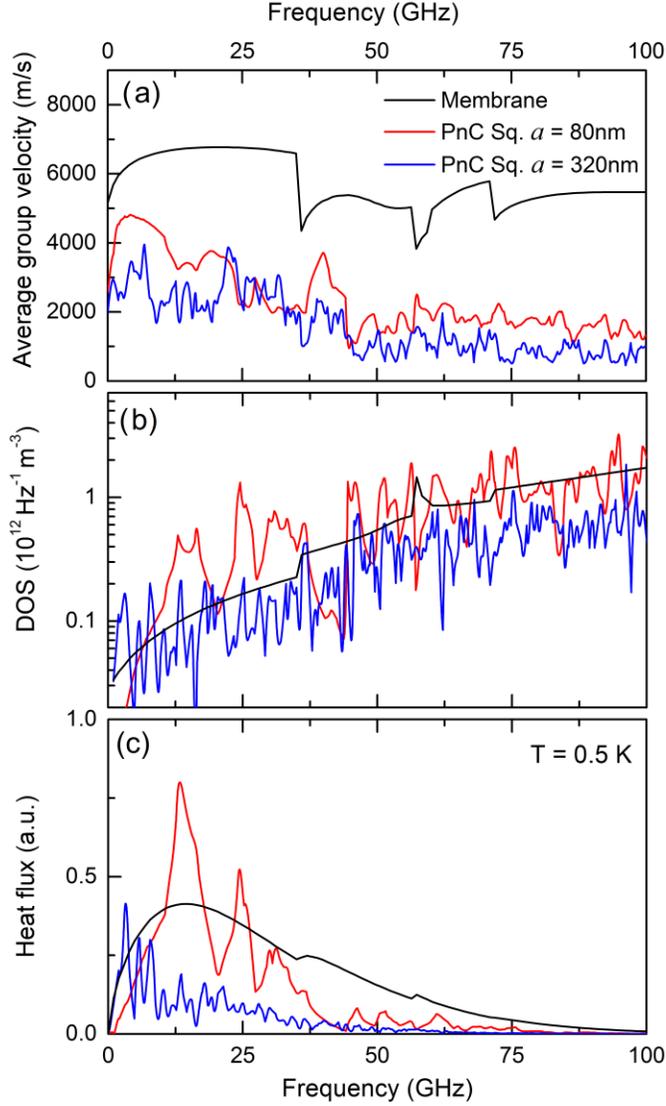

FIG. 3. (Color online) Spectra of (a) the average group velocity, (b) the DOS, and (c) the heat flux at 0.5 K, calculated for the membrane (black) and square lattice phononic structures with periods of $a = 80$ (red) and $a = 320$ nm (blue), and $r/a = 0.4$ for both structures. The thickness of all structures is $h = 80$ nm.

To compare quantitatively structures with different designs, we evaluate the thermal conductance of the structure as [11]:

$$G(T) \propto \frac{1}{2\pi^2} \sum_m \int_0^{FBZ} \hbar \omega_m \left| \frac{d\omega_m}{d\vec{k}} \right| \frac{\partial f(\omega,T)}{\partial T} d\vec{k} \qquad (3)$$

Note that this is a more correct approach than the one used in our previous work [13], where the thermal conductivity data is mistaken by 10 − 20%. In the present work we mostly consider relative thermal conductance given by $G_{PnC}$ /

$G_{Membrane}$ ratio, where $G_{PnC}$ and $G_{Membrane}$ are the thermal conductance of phononic crystals and a membrane, respectively. Figure 4 shows relative thermal conductance in phononic structures of different designs. All three lattices demonstrate decreasing trends as a function of the *r/a* ratio [Fig. 4(a)]. This fact is mostly explained by the reduction of group velocity due to the band flattening as the hole size is increased [12], though the reduction of DOS also plays a role. Therefore, the strongest reduction of thermal conductance takes place in the structures with the highest *r/a* ratio possible, since it is optimal for both coherent and incoherent [16,18,24] scattering mechanisms. This conclusion is also supported by recent molecular dynamics calculations [23].

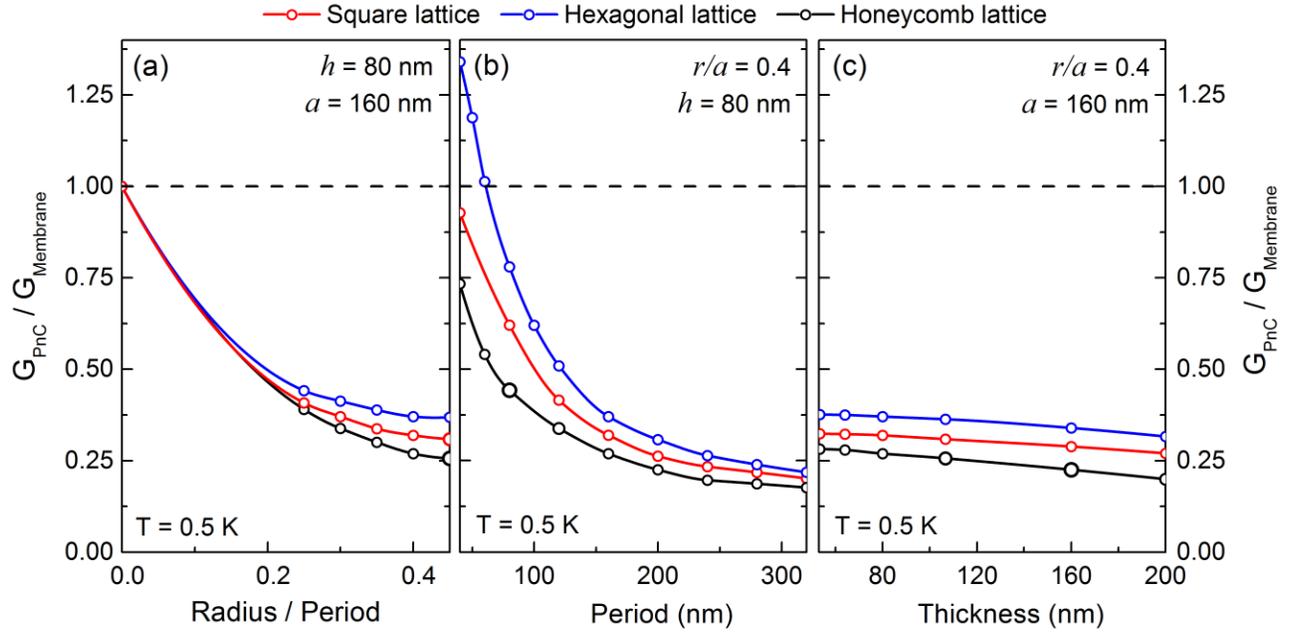

FIG. 4. (Color online) Relative thermal conductance in phononic structures with square (red), hexagonal (blue) and honeycomb (black) lattices, calculated at 0.5 K as (a) a function of *r/a* ratio, with constant thickness *h* = 80 and period *a* = 160 nm, (b) a function of the period with constant thickness *h* = 0.4 and period *a* = 160 nm, and (c) a function of the thickness with constant *r/a* = 0.4 and period *a* = 160 nm. Enlarged data points indicate structures with a phononic bandgap.

As a function of the period (with constant *r/a*), the reduction of thermal conductance is increased for the structures of all lattice types [Fig. 4(b)]. This dependence is explained mostly by the change of phonon density of states: increase of the period leads to a shift of the bands to lower frequencies. Such reduction of thermal conductance with the period is in agreement with recent experimental observations on 2D phononic crystals [11,34] as well as with experimental [35] and theoretical [36] works on 1D superlattices. An interesting consequence of this effect is that phononic structures with a relatively small period (*a* < 60 nm) can even demonstrate an enhancement of the thermal conductance, which is studied in detail in Ref. [13]. These data imply that in the presence of coherent scattering, a reduction of the period does not result necessarily in a more efficient phonon scattering, as expected from incoherent surface scattering due to an increase of the surface-to-volume ratio [31] or porosity [16,21].

Next, we study the dependence of thermal conductance on the thickness of the structures. For the unpatterned membranes this dependence has been investigated theoretically by Maasilta *et al*. [30,34,37]. Here we study the thickness dependence of the thermal conductance reduction by phononic crystals. This reduction is not changing significantly as the thickness is changed. The actual values of thermal conductance in both phononic crystals and membranes are increasing with thickness, but the $G_{PnC} / G_{Membrane}$ ratio remains approximately constant, decreasing only slightly as shown in Figure 4(c). This result is explained by the fact that while changes of period and radius-to-period ratio affect only phononic structures, change of thickness affects the dispersion of both phononic structure and unpatterned membrane by scaling the frequencies as $1/h$, while relative changes in the dispersion of phononic structure remain approximately the same. Therefore, an increase in thickness leads to only slight changes of the $G_{PnC} / G_{Membrane}$ ratio caused by the shift of modes to low frequency part and this consequently leads to an increase of the role of high frequency modes where the reduction of thermal conductance is stronger.

As far as different lattices are concerned, all three lattices demonstrate very similar trends of thermal conductance reduction. Yet, the thermal conductance in the structures with the same radius, period and thickness but different lattices seems to be lowest in honeycomb and highest in hexagonal structures, though as the period is increased this difference becomes negligible. This can be explained by the fact that a major difference in heat flux spectra appears at low frequencies, while high frequency parts are rather similar [Fig. 2(e)]. However, this finding implies that the choice of the lattice type can be made in order to achieve the best mechanical or electrical properties or the strongest incoherent surface scattering.

## IV. IMPACT OF TEMPERATURE

Since thermal conductance depends on the Bose–Einstein distribution, its reduction in phononic crystals can be affected by temperature, as more and more high-frequency bands become occupied at higher temperatures. Figure 5 shows the temperature dependence of the reduction of thermal conductance in phononic crystals as a function of temperature. At the low temperature limit, the thermal conductance of phononic crystals approaches, and, in case of a short period, even overcomes, that of the membrane (i.e. $G_{PnC} / G_{Membrane} > 1$). This reflects the thermal conductance boost effect [13]: at the temperatures about 0.1 K, only the first few bands are occupied, and these bands in phononic crystals have DOS higher than that in membrane [Fig 3(b)], while the reduction of group velocity in this region is moderate [Fig 3(a)]. However, as the temperature is increased, the reduction of thermal conductance strengthens for the structures of all periods. This is caused by an increasing impact of the bands in the high frequency range, where the band flattening is very strong, so the heat flux spectrum is reduced much stronger than at low temperatures, as shown in the inset of the Figure 5.

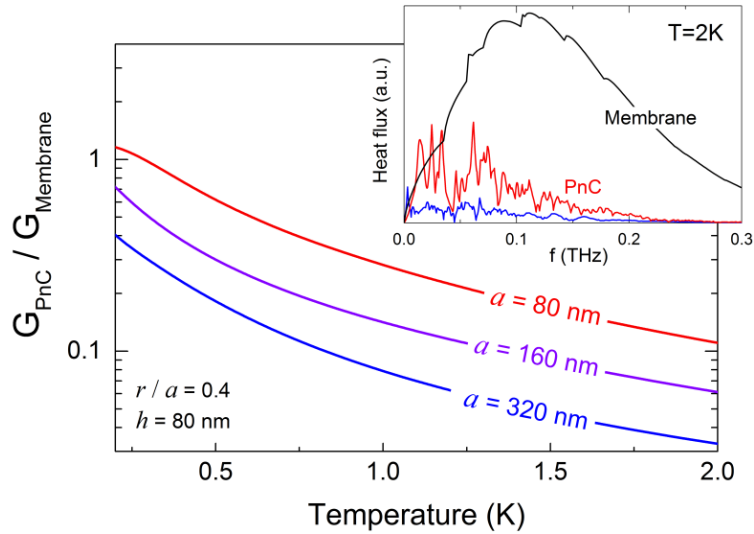

FIG. 5. (Color online) Relative thermal conductance in square lattice phononic structures ($h$ = 80 nm, $r / a$ = 0.4) with different periods ($a$ = 80, 160 and 320 nm) as a function of temperature. Inset shows the reduction of heat flux spectra in phononic crystals with $a$ = 80 and 320 nm as compared to unpatterned membrane.

On the other hand, the wave picture of phonons, discussed in this work, is relevant probably only at relatively low temperatures, when the thermal phonon wavelengths are comparable to the characteristic size of the structure [11,12,24]. Indeed, at room temperature, the phonon wavelengths are in the 1–5 nm range [25,38], which is often comparable to the surface roughness. But according to the formula of Ziman [39]: $p = \exp(-16 \pi^3 \delta^2 / \lambda^2)$, the wavelength ($\lambda$) should be at least of the order of 100 nm to achieve a specularity ($p$) of 0.5, for a surface roughness ($\delta$) of a few nanometers. So at room temperature the phonon boundary scattering is mostly incoherent (or diffusive) [24,25], and wave interference cannot develop. However, at low temperatures, thermal phonon wavelengths become longer, and coherent phonon scattering may play an important role. Zen et al. [11] demonstrated that at sub-kelvin temperatures it even can fully control the phonon transport, while Marconnet *et al.* [40] estimated that coherent scattering significantly impacts thermal conductivity in nanoscale structures only below 10 K. These estimations are in agreement with our recent experimental studies where signs of coherent scattering were found only at the temperatures below 7 K and reduction of thermal conductivity caused by this scattering was about 15% at 4 K [31,41]. Therefore, since on the one hand the reduction due to coherent scattering strengthens with temperature, while on the other hand coherent scattering itself becomes weaker, one may expect the existence of an optimal temperature (related to the surface roughness of a given structure) at which the reduction of thermal conductance reaches its maximum. This idea seems to in agreement with recent experimental observations [11,34], where the reduction of thermal conductance in phononic crystals strengthened according to the theoretical predictions only up to certain temperature (around 0.5 K) and then weakened again. Thus, the applications of coherent phonon control in phononic crystals are probably limited to low-temperature devices such as transition edge sensors [34].

## V. CONCLUSION

We have investigated theoretically the impact of structure design on the efficiency of coherent phonon scattering in 2D phononic crystals. Our results demonstrate that the reduction of thermal conductance by phononic nano-patterning becomes stronger as a function of the period and radius-to-period ratio, while the thickness of the structure has much weaker impact. As far as different lattice types are concerned, the structures of all three lattices demonstrate very similar values and trends of the reduction of thermal conductance, yet slightly stronger reduction is found in honeycomb lattice and the weakest in hexagonal lattice. Our study also demonstrates that the reduction of thermal conductance is temperature dependent and significantly strengthens with temperature, yet in realistic nanostructures coherent scattering plays an important role only at relatively low temperatures.

**ACKNOWLEDGMENTS**

This work was supported by the Project for Developing Innovation Systems of the Ministry of Education, Culture, Sports, Science and Technology (MEXT), Japan and by KAKENHI (25709090 and 15K13270).